\documentclass[twocolumn,showpacs,amsmath,amssymb]{revtex4}

\usepackage{amsmath} %
\usepackage{graphicx}
\usepackage{amsfonts}
\usepackage{dcolumn}
\usepackage{bm}

\begin{document}
\title{Inertial Effects on Kinetics of Motility-Induced Phase Separation}

\author{Jie Su}
\author{Huijun Jiang}
\thanks{E-mail: hjjiang3@ustc.edu.cn}
\author{Zhonghuai Hou}
\thanks{E-mail: hzhlj@ustc.edu.cn}
\affiliation{Department of Chemical Physics \& Hefei National Laboratory for Physical Sciences at Microscales, University of Science and Technology of China, Hefei, Anhui 230026, China}
\date{\today}

\begin{abstract}
Motility-induced phase separation (MIPS) is of great importance and has been extensively researched in overdamped systems, nevertheless, what impacts inertia will bring on kinetics of MIPS is lack of investigation. Here, we find that, not only the phase transition changes from continuous to discontinuous, but also the formation of clusters exhibits a nucleation-like process without any coarsening regime, different from spinodal decomposition in the overdamped case. This remarkable kinetics stems from a competition between activity-induced accumulation of particles and inertia-induced suppression of clustering process. More interestingly, the discontinuity of MIPS still exists even when the ratio of particle mass to the friction coefficient reduces to be very small such as $10^{-4}$. Our findings emphasize the importance of inertia in kinetics of MIPS, and may open a new perspective on understanding the nature of MIPS in active systems.
\end{abstract}

\maketitle
Active systems transducing energy into systematic movement to drive themselves far from equilibrium\cite{a1} have attracted tremendous interests and become one of the hottest research topics across physical, chemical, materials and biological sciences. In comparison with their passive counterparts, active systems exhibit many novel nonequilibrium behaviors, such as emergence of dynamic chirality\cite{a4,a5,a6,a18}, polar swarms\cite{a8,a7,a11}, collective vortex\cite{a7,a11,a2,a3}, and particularly motility-induced phase separation (MIPS)\cite{MIPS1,MIPS2,MIPS3,MIPS4,MIPS5,MIPS6,MIPS7,MIPS8,MIPS9,MIPS12,MIPS13,MIPS14,MIPS10,MIPS11,MIPS15,MIPS16,MIPS17,MIPS18,MIPS19,MIPS20,MIPS21,MIPS22,MIPS23,MIPS24}. MIPS was first introduced by Cates \emph{et al}. in systems of run-and-tumble particles\cite{MIPS1}, and then in systems consisting of active brownian particles (ABPs) which are purely repulsive without any attractive interactions\cite{MIPS2}. Generally, active particles tend to accumulate where they move more slowly and will slow down at high density for steric reasons, which then creates positive feedback leading to MIPS between dense and dilute fluid phases\cite{MIPS3}. In order to provide further deep insight in the MIPS process, Hagan \emph{et al}. investigated the phase separation kinetics in overdamped ABP systems by numerical simulations\cite{MIPS2}. They found that active systems quenched close to the binodal exhibit a nucleation-like behavior characterized by a discontinuous transition from a single phase to MIPS, while ones quenched more deeply across the spinodal undergo spinodal decomposition with a coarsening regime. In addition, Such a physical picture was supported by various theoretical analysis of MIPS based on the overdamped ABP model, such as the kinetic model\cite{MIPS2,MIPS5}, effective Cahn-Hilliard equation\cite{MIPS6,MIPS7,MIPS8,MIPS9}, and swim pressure\cite{MIPS12,MIPS13,MIPS14}.

Notice that, most of previous works were based on the overdamped Brownian model. The overdamped approximation works well in many novel nonequilibrium behaviors in active systems. However, this approximation has also been found to encounter several difficulties in some applications due to inertial effects\cite{inertia12,inertia13,inertia14,inertia10,inertia1,inertia11,inertia4,inertia5,inertia6,inertia7,inertia8,inertia9,inertia3,inertia2}. Especially, inertia-induced hidden entropy production has been revealed even in the limit of small inertia\cite{inertia12,inertia13,inertia14}. Besides, inertia-induced coexistence of different kinetic temperatures has been reported when the reduced mass measuring the impact of inertia changes from $10^{-1}$ to $10^{-4}$\cite{inertia10}. It is then very intriguing to understand how inertia affects the phase separation kinetics of the well-known MIPS.

Motivated by this, we revealed inertial effects on kinetics of MIPS by investigating activity-dependent steady-state local density distributions (LDDs) and time-dependent cluster-growth kinetics (TDGK) for systems with and without inertia. For overdamped systems undergoing spinodal decomposition, the phase transition is continuous characterized by the continuously changing peaks of steady-state LDDs and a coarsening growth of clusters. When inertia presents, the phase transition changes to be discontinuous, where a jump of LDD peaks is found and the formation of clusters exhibits a nucleation-like process. The inertia-induced change of MIPS kinetics is further validated by the enhanced discontinuity of steady-state LDD peaks and nucleation barrier for cluster growth as inertia increases. Detailed analysis reveals that, inertia will change the movement of active particles when they collide with each other, which leads to inertia-induced suppression of clustering process. The competition between activity-induced accumulation and inertia-induced suppression eventually results in the distinct phase separation kinetics. More interestingly, the discontinuity of MIPS still exists even when the ratio of particle mass to the friction coefficient reduces to be very small such as $10^{-4}$, implying that the MIPS kinetics of underdamped systems in the limit of small inertia is totally different from that of overdamped systems and some hidden information may be lost when the overdamped approximation is applied.

We consider a quasi two-dimensional system with size $L$ and periodic boundary conditions consisting of $N$ active spherical particles with diameter $\sigma$. For the situation in absence of inertial effects, the motion of each ABP obeys the following overdamped Langevin equations:

\begin{equation}
\gamma\frac{d\mathbf{r}_i}{dt}=\gamma v_0\mathbf{n}_i-\sum_{j=1}^N\frac{\partial{U(\mathbf{r}_{ij})}}{\partial{\mathbf{r}_i}} +\mathbf{f}_i,\label{eq:translation1}
\end{equation}
\begin{equation}
\gamma_r\frac{d\phi_i}{dt}=g_i.\label{eq:rotation1}
\end{equation}

\noindent Herein, $\gamma$ is the friction coefficient, $\mathbf{r}_{i}$ represents the $ith$ ABP's position, $\mathbf{r}_{ij}=\mathbf{r}_{i}-\mathbf{r}_{j}$, $v_0$ and $\mathbf{n}_i=(\cos(\phi_i),\sin(\phi_i))$ denote respectively the amplitude and direction of active velocity for the $ith$ ABP with $\phi_i$ the angle of $\mathbf{n}_i$. The exclusive-volume effect between a pair of ABPs is described by the WCA potential: $U(\mathbf{r}_{ij})=4\epsilon[(\frac{\sigma}{r_{ij}})^{12}-(\frac{\sigma}{r_{ij}})^6+\frac{1}{4}]$ for $r_{ij}<2^{1/6}\sigma$, and $U=0$ otherwise, with $\epsilon$ the interaction strength and $r_{ij}$ the norm of $\mathbf{r}_{ij}$. $\mathbf{f}_i$ denotes the random force satisfying the fluctuation-dissipation relation $\langle\mathbf{f}_i(t)\mathbf{f}_j(t')\rangle=2k_BT/\gamma\delta_{ij}\delta(t-t')$, where $k_BT$ is an effective thermal energy quantifying the noise strength. In the rotational equation (Eqs. (\ref{eq:rotation1})), $\gamma_r=\sigma^2\gamma/3$ is the rotational friction coefficient, $g_i$ is the rotational fluctuation satisfying $\langle g_i(t)g_j(t')\rangle=2k_BT/\gamma_r\delta_{ij}\delta(t-t')$.

When inertia is taken into account, the motion of each active inertial particle (AIP) can be described as the following underdamped Langevin equations:

\begin{equation}
M\frac{d^2\mathbf{r}_i}{dt^2}+\gamma\frac{d\mathbf{r}_i}{dt}=f_0\mathbf{n}_i-\sum_{j=1}^N\frac{\partial{U(\mathbf{r}_{ij})}}{\partial{\mathbf{r}_i}} +\mathbf{f}_i,\label{eq:translation2}
\end{equation}
\begin{equation}
J\frac{d^2\phi_i}{dt^2}+\gamma_r\frac{d\phi_i}{dt}=g_i.\label{eq:rotation2}
\end{equation}

\noindent Herein, $M$ is the mass of AIPs, $J=\sigma^2M/10$ denotes the moment of inertia, and $f_0$ represents active force of each AIP. In the limit of $M\rightarrow0$, these underdamped equations recover to the overdamped ones Eqs.(\ref{eq:translation1}) and (\ref{eq:rotation1}).

In simulations, parameters are dimensionless by $\sigma$, $\gamma$ and $k_BT$, so that the basic unit for time is $\gamma\sigma^2/(k_BT)$. We fix $M=0.015$, $L=200$, $N=30720$, $\epsilon=k_BT$, if not otherwise stated. Thereby the averaged number density of each system is $\rho_0=N/L^2=0.768$. For consistency, all of the following results are obtained from simulations of $2\times10^7$ steps with the time step $\Delta t=10^{-5}$ and random initial conditions.
According to that in literature, the criterion of MIPS is the steady-state LDD changes from an unimodal distribution (corresponding to a single phase with one dominant density) to a bimodal one (indicating the coexistence of two phases with two different dominant densities)\cite{MIPS2,inertia11}.

\begin{figure}
\begin{centering}
\includegraphics[width=1.0\columnwidth]{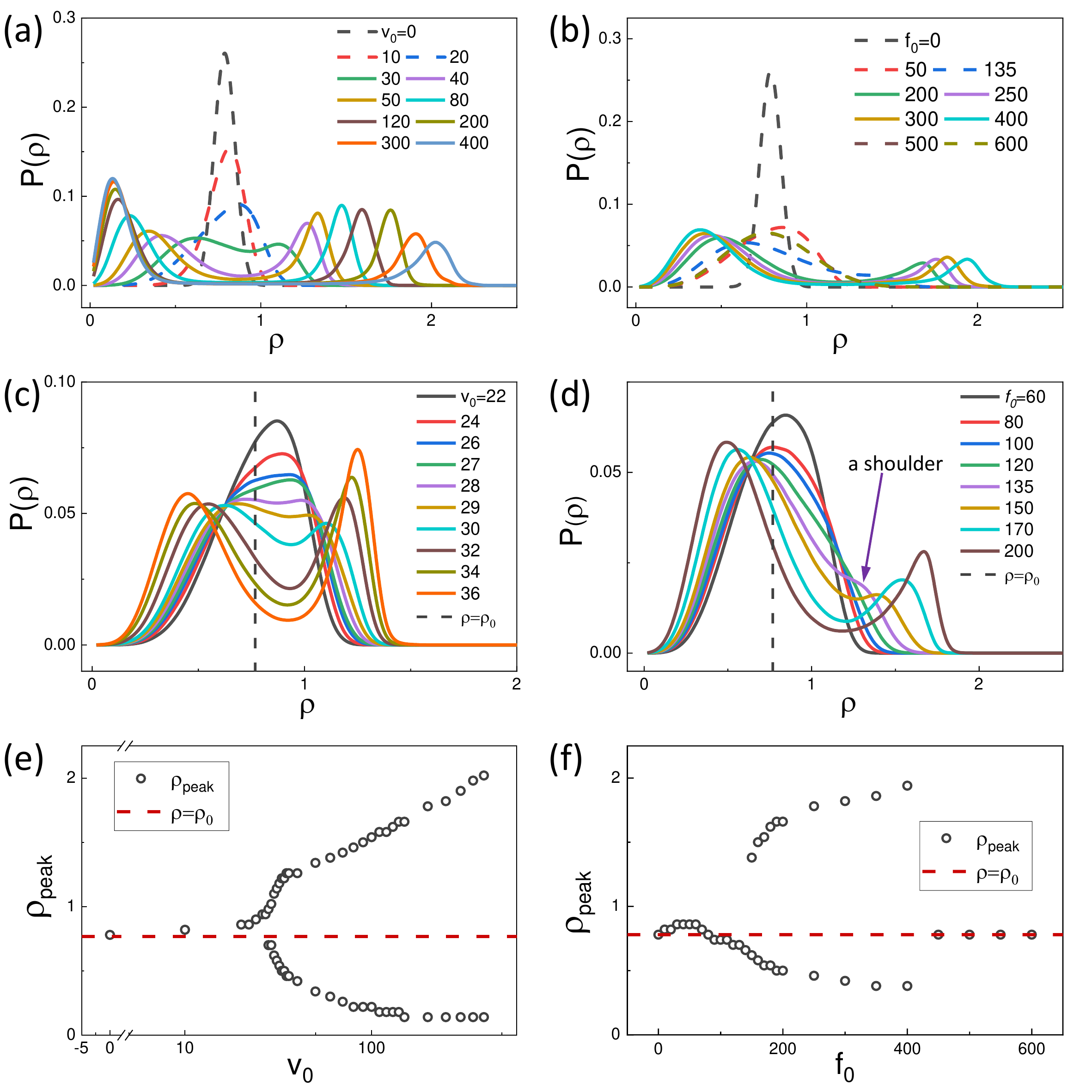}
\par\end{centering}
\caption{Steady-state LDDs for various activities in (a) the ABP system, and (b) the AIP system. The dashed lines are unimodal (non-MIPS) and solid ones are bimodal (MIPS). Steady-state LDDs near the transition threshold are zoomed-in for (c) the ABP system and (d) the AIP system. Dependence of $\rho_{peak}$ (position of the steady-state LDD peak) as functions of activities for (e) the ABP system, and (f) the AIP system. The dashed lines denote the averaged number density $\rho=\rho_0$.}
\label{fig:result1}
\end{figure}

Firstly, we focus on the steady-state LDDs of MIPS in ABP or AIP systems for varying activity. For the ABP system, it is observed that the steady-state LDD changes from unimodal to binodal when activity increases across a threshold $v_0^*$ around $v_0=30$ as shown in Fig.~\ref{fig:result1} (a), in alignment with MIPS reported in previous literature\cite{MIPS2,MIPS3}. For the AIP system, the steady-state LDD also changes from unimodal to binodal as activity increases across a threshold $f_0^*$ around $f_0=150$, but it will go back to be unimodal again when activity further increases much larger such as $f_0>400$ as shown in Fig.~\ref{fig:result1} (b), indicating that the AIP system influenced by inertia reenters into a single phase, which has also been observed by L\"{o}wen \emph{et al}.\cite{inertia10}. To take a closer look at the phase separation kinetics, steady-state LDDs around the threshold are then plotted in Fig.~\ref{fig:result1} (c) and (d). In the ABP system (Fig.~\ref{fig:result1} (c)), position of the steady-state LDD peak keeps almost unchanged before the threshold $v_0^*=28$, while the peak divides into two peaks with almost the same shape just after the threshold. Interestingly, a brand new type of activity-dependent steady-state LDD transition is found in the AIP system. As shown in Fig.~\ref{fig:result1} (d), position of the steady-state LDD peak shifts to low densities before the threshold, and the peak of the dense phase arises at a rather higher density in comparison with the one in the ABP system after the threshold. Taking $f_0=135$ as an example, an obvious shoulder can be found near $\rho=1.25$ in Fig.~\ref{fig:result1} (d), ready to form a peak at high density.

To quantitatively describe the phase separation kinetics, we plot dependence of $\rho_{peak}$ (position of the steady-state LDD peak) as functions of activity for ABPs and AIPs in Fig.~\ref{fig:result1} (e) and (f), respectively. In the ABP system (Fig.~\ref{fig:result1} (e)), $\rho_{peak}$ divides into two parts continuously after $v_0>=v_0^*$, indicating that MIPS occurs via spinodal decomposition in the overdamped case, which is in consistence with that reported previously\cite{MIPS2,MIPS3,MIPS6}. In contrast, for the AIP system (Fig.~\ref{fig:result1} (f)), $\rho_{peak}$ of the high density phase appears discontinuously with a sharp jump ($f_0=150$ in Fig.~\ref{fig:result1} (f)), representing that MIPS is nucleation-like. In short, inertia can strongly affect the nature of MIPS by changing the transition from continuous to discontinuous.

\begin{figure}
\begin{centering}
\includegraphics[width=1.0\columnwidth]{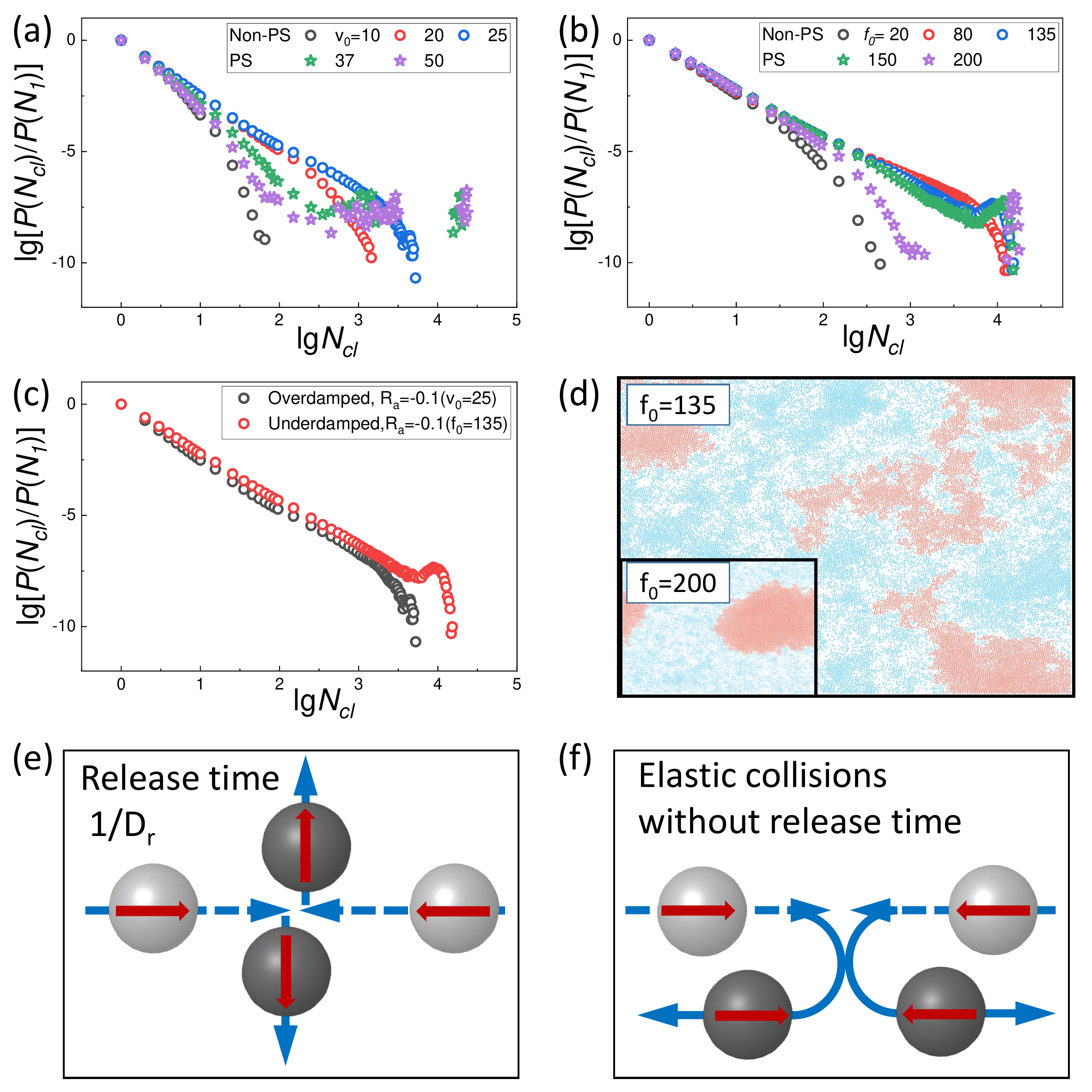}
\par\end{centering}
\caption{The mechanism for the inertia-influenced phase separation kinetics. CSDs on varying activities for (a) the ABP system, and (b) the AIP system. (c) A comparison of CSDs between ABP and AIP systems with the same distance $R_a=-0.1$ to the transition threshold. (d) A snapshot for the steady-state AIP system at $f_0=135$ just before the transition threshold. The biggest loosened cluster is highlighted by red, in relative to the densely packed cluster at $f_0=200$ after the transition threshold in the inset. Sketches of collision events (e) without or (f) with inertial effects. Light or dark balls represent active particles before or after collisions, respectively. Red arrows are directions of active forces and the blue ones denote particle trajectories.}
\label{fig:result2}
\end{figure}

In order to figure out the underlying mechanism for inertial effects on phase separation kinetics, we plot cluster size distributions (CSDs) for varying activities in ABP and AIP systems in Fig.~\ref{fig:result2} (a) and (b), respectively. Here, $P(N_{cl})$ for the cluster of size $N_{cl}$ is rescaled by $P(N_1)$\cite{MIPS5,MIPS20}. It can be observed that CSD changes first from exponential to power-law then to bimodal as activity increases in both ABP and AIP systems. However, the power-law distributions near transition thresholds seem to be different for these two systems. In order to be comparable for CSDs of ABP and AIP systems, we define $R_a=(f_0^{se}-f_0^*)/f_0^*=(v_0^{se}-v_0^*)/v_0^*$ to measure the distance of the present activities to transition thresholds. CSDs for $R_a=-0.1$ just before the threshold (i.e., $v_0=25$ and $f_0=135$) are presented in Fig.~\ref{fig:result2} (c). Interestingly, CSD is power-law with an exponential tail for the ABP system, while power-law with a small peak arising at large $N_{cl}$ for the AIP system. This small peak represents that a loosened cluster with relatively large size appears, which can be demonstrated by the snapshot for the AIP system with $f_0=135$ as plotted in Fig.~\ref{fig:result2} (d). It is noted that the steady-state LDD for $f_0=135$ still keeps unimodal with a shoulder at large density as shown in Fig.~\ref{fig:result1}(d). The delayed separation of density relative to separation of cluster size thus leads to the inertia-induced discontinuous transition observed in Fig.~\ref{fig:result1}.

The delayed separation can be related to inertial effects similar to that reported by L\"{o}wen \emph{et al}.\cite{inertia1}. As sketched in Fig.~\ref{fig:result2} (e), when two ABPs collide head-on with each other, they will stop and wait until their propulsion directions separated from each other, leading to a released time $1/D_r$ with $D_r$ the rotational diffusion coefficient. If other particles approach the particle pair within such a released time, the pair will be surrounded by more particles, leading to accumulation of particles to form clusters, further resulting in MIPS\cite{MIPS3,inertia1}. However, when two AIPs collide, they will bounce back directly without any released time due to inertia-induced elastic collision (Fig.~\ref{fig:result2} (f)). This elastic collision induced by inertia will increase the averaged distance among AIPs in clusters, leading to a lower cluster density in comparison with that when inertia is absent. In general, activity will induce accumulation of particles to form clusters, while inertia suppresses the clustering process. The competition between these two factors then results in the distinct kinetics of MIPS in the AIP system.

\begin{figure}
\begin{centering}
\includegraphics[width=1.0\columnwidth]{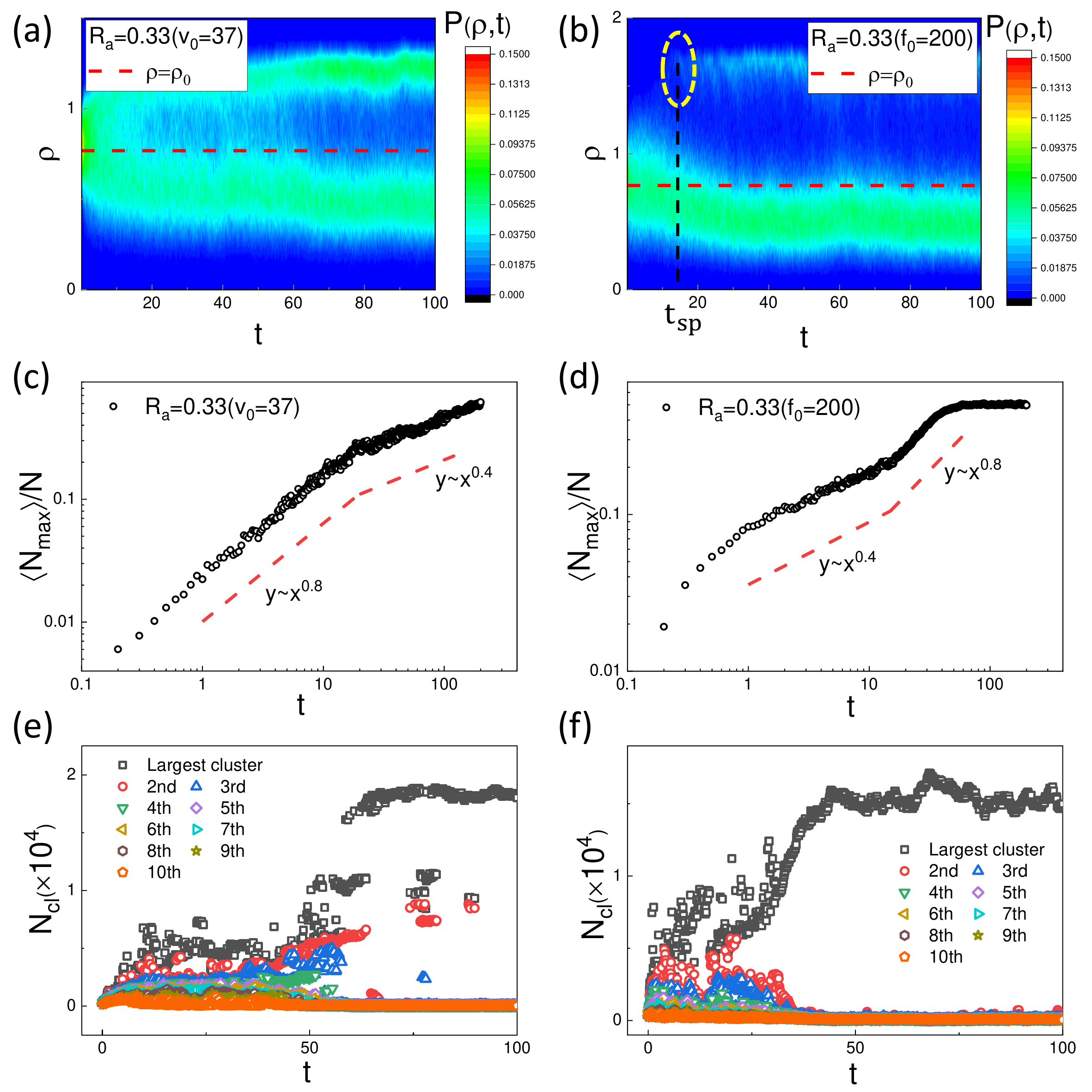}
\par\end{centering}
\caption{Time-dependent LDDs in $t$-$\rho$ plane located in MIPS regions for (a) the ABP system ($v_0=37$, $R_a=0.33$), and (b) the AIP system ($f_0=200$, $R_a=0.33$). A discontinuous jump appearing at the separation time ($t_{sp}$) can be observed in the AIP system. Dependence of the ensemble-averaged largest cluster size on time for (c) the ABP system and (d) the AIP system. Time series of ten largest clusters for (e) the ABP system ($v_0=37$), and (f) the AIP system ($f_0=200$).
}
\label{fig:result4}
\end{figure}

Such a mechanism can further be verified by time-dependent cluster-growth kinetics (TDGK) for ABP and AIP systems. It is known that, in the ABP system, only the activity-induced accumulation of particles works, leading to a TDGK of spinodal-decomposition type with a fast growth of small clusters at early stage and a slow cluster-coarsening process later\cite{MIPS2,MIPS10}. It is expected that TDGK should be different when inertia-induced suppression of clustering process takes place. In Fig.~\ref{fig:result4} (a) and (b), time-dependent LDDs in ABP and AIP systems for $R_a=0.33$ are plotted, respectively. Similar to steady-state LDDs, time-dependent LDDs change from unimodal to bimodal continuously for the ABP system, while a discontinuous jump occurs at a separation time ($t_{sp}$) for the AIP system. In Fig.~\ref{fig:result4} (c) and (d), time evolution of the ensemble-averaged largest cluster size $\langle N_{max}\rangle$ normalised by the total particle number $N$ for ABP and AIP systems are plotted, respectively. For the ABP system (Fig.~\ref{fig:result4} (c)), the function of $\langle N_{max}\rangle/N$ on $t$ in the double logarithmic coordinate shows two different slopes, i.e., around $0.8$ at the early stage and about $0.4$ later, in accordance with the aforementioned TDGK for the ABP system. For the AIP system, the function of $\langle N_{max}\rangle/N$ on $t$ in the double logarithmic coordinate can also be divided into two part as shown in Fig.~\ref{fig:result4} (d). Different from the ABP system, the slope is smaller than $0.4$ at the early stage and reaches about $0.8$ later, indicating that TDGK influenced by inertia is nucleation-like with a slow nucleation process firstly followed by a fast cluster-growth.

Such a growth kinetics can further be supported by time-dependent pathways of cluster growth. Time series of the ten largest clusters in ABP and AIP systems for $R_a=0.33$ are presented in Fig.~\ref{fig:result4} (e) and (f), respectively. It is observed that cluster sizes change frequently by sharp jumps in the ABP system (Fig.~\ref{fig:result4} (e)), indicating that TDGK is of cluster-coarsening type. In the AIP system, however, change of cluster sizes is more continuous, implying that TDGK is of a cluster-growth type(Fig.~\ref{fig:result4} (f)). Movies S1 and S2 for the different growth processes can be found in supplemental materials. In short, inertia can also affect the nature of MIPS by changing the time-dependent cluster-growth kinetics from typical spinodal decomposition to a nucleation-like type.

\begin{figure}
\begin{centering}
\includegraphics[width=1.0\columnwidth]{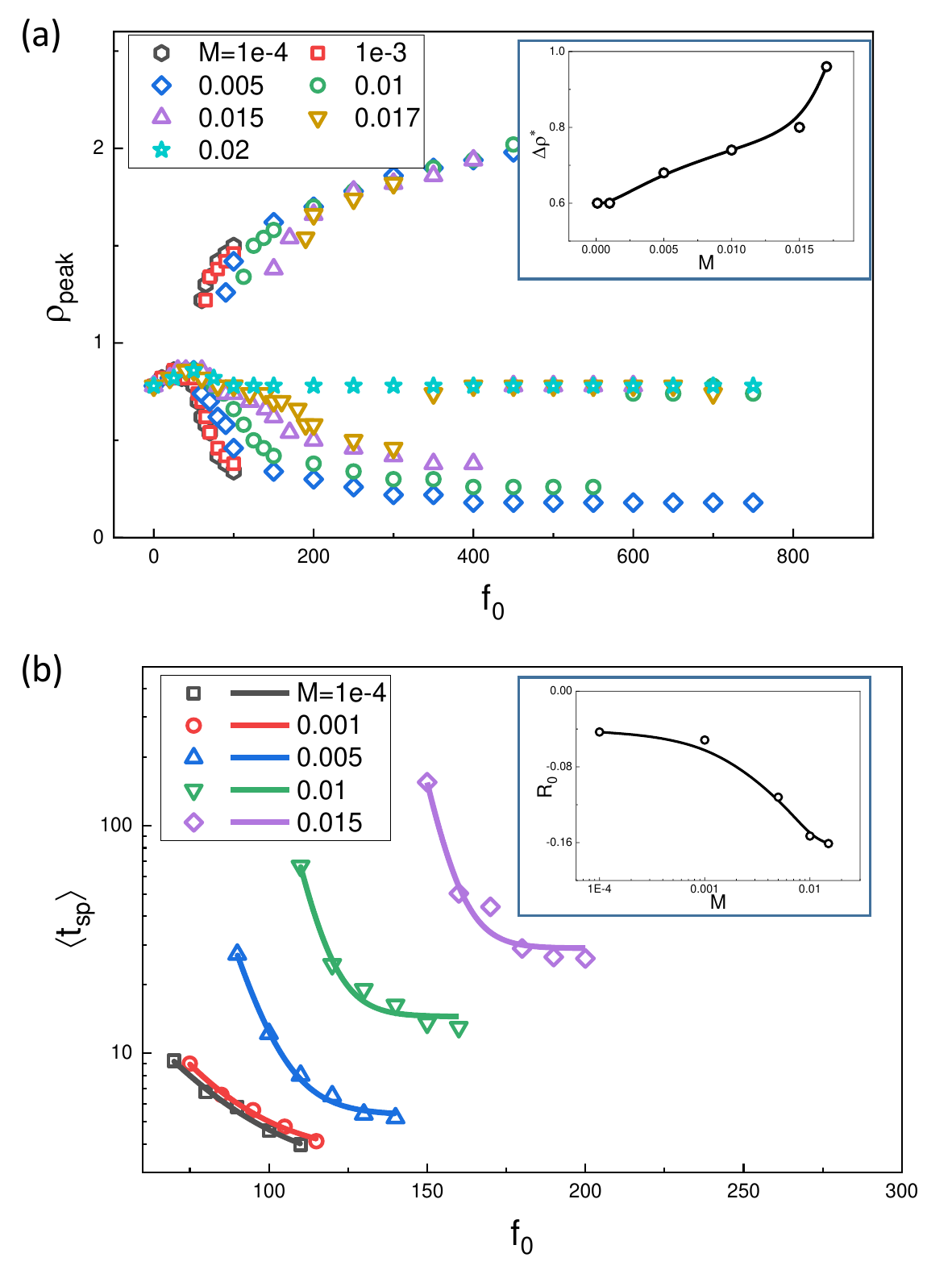}
\par\end{centering}
\caption{Phase separation kinetics influenced by different inertia. (a) Dependence of $\rho_{peak}$ as functions of $f_0$ in the AIP system for varying $M$. The inset is dependence of the discontinuous jump size $\Delta\rho^{*}$ as a function of $M$. (b) Dependence of the ensemble-averaged separation time $\langle t_{sp}\rangle$ as functions of $f_0$ in the AIP system for varying $M$. Curves are the exponential fitting for each group of data. The inset is dependence of the exponent $R_0$ as a function of $M$.}
\label{fig:result5}
\end{figure}

Furthermore, we want to know how the revealed change of phase separation kinetics depends on different inertia. Intensive simulations for AIPs with different particle mass $M$ are performed. For steady-state LDDs, dependence of $\rho_{peak}$ as functions of $f_0$ for varying $M$ is plotted in Fig.~\ref{fig:result5} (a). The discontinuous transition can always be observed as $M$ increases, until $M$ is large enough such as $M=0.2$ where no MIPS occurs as reported by L\"{o}wen \emph{et al}.\cite{inertia10}. Size of the discontinuous jump $\Delta\rho^{*}$ measured by the difference between $\rho_{peak}$s of the dense and dilute phases for different $M$ is plotted in the inset of Fig.~\ref{fig:result5} (a). Clearly, larger inertia will lead to larger $\Delta\rho^{*}$. Ensemble-averaged separation time $\langle t_{sp}\rangle$ for the time-dependent cluster-growth kinetics in the AIP system with different $M$ is shown in Fig.~\ref{fig:result5} (b). It is observed that $\langle t_{sp}\rangle$ exponentially decreases as activity $f_0$ increases, demonstrating again that the formation of clusters with inertial effects is a nucleation-like process. As presented in the inset of Fig.~\ref{fig:result5} (b), the absolute value of the exponent $R_0$ which can be related to the nucleation barrier gets larger with the increase of $M$, indicating that larger inertia will increase the nucleation barrier, further leading to a stronger suppression of MIPS. More interestingly, both the discontinuous jump and the nucleation barrier can still be observed for $M/\gamma=10^{-4}$, and tends to keep even for $M/\gamma\rightarrow0$, implying that the MIPS kinetics of underdamped systems in the limit of small inertia is totally different from that of overdamped systems.

In summary, inertial effects on kinetics of motility-induced phase separation have been investigated. Besides activity-induced accumulation of particles, inertia-induced suppression of clustering process was found when inertia presents. The competition between these two factors affects strongly the nature of MIPS by changing the transition from continuous to discontinuous and changing the time-dependent cluster-growth kinetics from typical spinodal decomposition to a nucleation-like type. Such effects can be observed even when the ratio of particle mass to the friction coefficient reduces to $10^{-4}$, implying that the MIPS kinetics of underdamped systems in the limit of small inertia is totally different from that of overdamped systems and some hidden information may be lost when the overdamped approximation is applied. Our findings emphasize the importance of inertia in kinetics of MIPS, and may open a new perspective on understanding the nature of MIPS in active systems.

This work is supported by MOST(2016YFA0400904, 2018YFA0208702), NSFC (21973085, 21833007, 21790350, 21673212, 21521001, 21473165), and Anhui Initiative in Quantum Information Technologies (AHY090200).


\begin{thebibliography}{48}
\expandafter\ifx\csname natexlab\endcsname\relax\def\natexlab#1{#1}\fi
\expandafter\ifx\csname bibnamefont\endcsname\relax
  \def\bibnamefont#1{#1}\fi
\expandafter\ifx\csname bibfnamefont\endcsname\relax
  \def\bibfnamefont#1{#1}\fi
\expandafter\ifx\csname citenamefont\endcsname\relax
  \def\citenamefont#1{#1}\fi
\expandafter\ifx\csname url\endcsname\relax
  \def\url#1{\texttt{#1}}\fi
\expandafter\ifx\csname urlprefix\endcsname\relax\def\urlprefix{URL }\fi
\providecommand{\bibinfo}[2]{#2}
\providecommand{\eprint}[2][]{\url{#2}}

\bibitem[{\citenamefont{Ramaswamy}(2010)}]{a1}
\bibinfo{author}{\bibfnamefont{S.}~\bibnamefont{Ramaswamy}},
  \bibinfo{journal}{Annu. Rev. Condens. Matter Phys.}
  \textbf{\bibinfo{volume}{1}}, \bibinfo{pages}{323} (\bibinfo{year}{2010}).

\bibitem[{\citenamefont{Riedel et~al.}(2005)\citenamefont{Riedel, Kruse, and
  Howard}}]{a4}
\bibinfo{author}{\bibfnamefont{I.~H.} \bibnamefont{Riedel}},
  \bibinfo{author}{\bibfnamefont{K.}~\bibnamefont{Kruse}}, \bibnamefont{and}
  \bibinfo{author}{\bibfnamefont{J.}~\bibnamefont{Howard}},
  \bibinfo{journal}{Science} \textbf{\bibinfo{volume}{309}},
  \bibinfo{pages}{300} (\bibinfo{year}{2005}).

\bibitem[{\citenamefont{DiLuzio et~al.}(2005)\citenamefont{DiLuzio, Turner,
  Mayer, Garstecki, Weibel, Berg, and Whitesides}}]{a5}
\bibinfo{author}{\bibfnamefont{W.~R.} \bibnamefont{DiLuzio}},
  \bibinfo{author}{\bibfnamefont{L.}~\bibnamefont{Turner}},
  \bibinfo{author}{\bibfnamefont{M.}~\bibnamefont{Mayer}},
  \bibinfo{author}{\bibfnamefont{P.}~\bibnamefont{Garstecki}},
  \bibinfo{author}{\bibfnamefont{D.~B.} \bibnamefont{Weibel}},
  \bibinfo{author}{\bibfnamefont{H.~C.} \bibnamefont{Berg}}, \bibnamefont{and}
  \bibinfo{author}{\bibfnamefont{G.~M.} \bibnamefont{Whitesides}},
  \bibinfo{journal}{Nature} \textbf{\bibinfo{volume}{435}},
  \bibinfo{pages}{1271} (\bibinfo{year}{2005}).

\bibitem[{\citenamefont{K{\"u}mmel et~al.}(2013)\citenamefont{K{\"u}mmel, ten
  Hagen, Wittkowski, Buttinoni, Eichhorn, Volpe, L{\"o}wen, and
  Bechinger}}]{a6}
\bibinfo{author}{\bibfnamefont{F.}~\bibnamefont{K{\"u}mmel}},
  \bibinfo{author}{\bibfnamefont{B.}~\bibnamefont{ten Hagen}},
  \bibinfo{author}{\bibfnamefont{R.}~\bibnamefont{Wittkowski}},
  \bibinfo{author}{\bibfnamefont{I.}~\bibnamefont{Buttinoni}},
  \bibinfo{author}{\bibfnamefont{R.}~\bibnamefont{Eichhorn}},
  \bibinfo{author}{\bibfnamefont{G.}~\bibnamefont{Volpe}},
  \bibinfo{author}{\bibfnamefont{H.}~\bibnamefont{L{\"o}wen}},
  \bibnamefont{and}
  \bibinfo{author}{\bibfnamefont{C.}~\bibnamefont{Bechinger}},
  \bibinfo{journal}{Phys. Rev. Lett.} \textbf{\bibinfo{volume}{110}},
  \bibinfo{pages}{198302} (\bibinfo{year}{2013}).

\bibitem[{\citenamefont{Su et~al.}(2019)\citenamefont{Su, Jiang, and
  Hou}}]{a18}
\bibinfo{author}{\bibfnamefont{J.}~\bibnamefont{Su}},
  \bibinfo{author}{\bibfnamefont{H.}~\bibnamefont{Jiang}}, \bibnamefont{and}
  \bibinfo{author}{\bibfnamefont{Z.}~\bibnamefont{Hou}}, \bibinfo{journal}{Soft
  Matter} \textbf{\bibinfo{volume}{15}}, \bibinfo{pages}{6830}
  (\bibinfo{year}{2019}).

\bibitem[{\citenamefont{Bricard et~al.}(2013)\citenamefont{Bricard, Caussin,
  Desreumaux, Dauchot, and Bartolo}}]{a8}
\bibinfo{author}{\bibfnamefont{A.}~\bibnamefont{Bricard}},
  \bibinfo{author}{\bibfnamefont{J.-B.} \bibnamefont{Caussin}},
  \bibinfo{author}{\bibfnamefont{N.}~\bibnamefont{Desreumaux}},
  \bibinfo{author}{\bibfnamefont{O.}~\bibnamefont{Dauchot}}, \bibnamefont{and}
  \bibinfo{author}{\bibfnamefont{D.}~\bibnamefont{Bartolo}},
  \bibinfo{journal}{Nature} \textbf{\bibinfo{volume}{503}}, \bibinfo{pages}{95}
  (\bibinfo{year}{2013}).

\bibitem[{\citenamefont{Yan et~al.}(2016)\citenamefont{Yan, Han, Zhang, Xu,
  Luijten, and Granick}}]{a7}
\bibinfo{author}{\bibfnamefont{J.}~\bibnamefont{Yan}},
  \bibinfo{author}{\bibfnamefont{M.}~\bibnamefont{Han}},
  \bibinfo{author}{\bibfnamefont{J.}~\bibnamefont{Zhang}},
  \bibinfo{author}{\bibfnamefont{C.}~\bibnamefont{Xu}},
  \bibinfo{author}{\bibfnamefont{E.}~\bibnamefont{Luijten}}, \bibnamefont{and}
  \bibinfo{author}{\bibfnamefont{S.}~\bibnamefont{Granick}},
  \bibinfo{journal}{Nat. Mater.} \textbf{\bibinfo{volume}{15}},
  \bibinfo{pages}{1095} (\bibinfo{year}{2016}).

\bibitem[{\citenamefont{Karani et~al.}(2019)\citenamefont{Karani, Pradillo, and
  Vlahovska}}]{a11}
\bibinfo{author}{\bibfnamefont{H.}~\bibnamefont{Karani}},
  \bibinfo{author}{\bibfnamefont{G.~E.} \bibnamefont{Pradillo}},
  \bibnamefont{and} \bibinfo{author}{\bibfnamefont{P.~M.}
  \bibnamefont{Vlahovska}}, \bibinfo{journal}{Phys. Rev. Lett.}
  \textbf{\bibinfo{volume}{123}}, \bibinfo{pages}{208002}
  (\bibinfo{year}{2019}).

\bibitem[{\citenamefont{Sumino et~al.}(2012)\citenamefont{Sumino, Nagai,
  Shitaka, Tanaka, Yoshikawa, Chat{\'e}, and Oiwa}}]{a2}
\bibinfo{author}{\bibfnamefont{Y.}~\bibnamefont{Sumino}},
  \bibinfo{author}{\bibfnamefont{K.~H.} \bibnamefont{Nagai}},
  \bibinfo{author}{\bibfnamefont{Y.}~\bibnamefont{Shitaka}},
  \bibinfo{author}{\bibfnamefont{D.}~\bibnamefont{Tanaka}},
  \bibinfo{author}{\bibfnamefont{K.}~\bibnamefont{Yoshikawa}},
  \bibinfo{author}{\bibfnamefont{H.}~\bibnamefont{Chat{\'e}}},
  \bibnamefont{and} \bibinfo{author}{\bibfnamefont{K.}~\bibnamefont{Oiwa}},
  \bibinfo{journal}{Nature} \textbf{\bibinfo{volume}{483}},
  \bibinfo{pages}{448} (\bibinfo{year}{2012}).

\bibitem[{\citenamefont{Jiang et~al.}(2017)\citenamefont{Jiang, Ding, Pu, and
  Hou}}]{a3}
\bibinfo{author}{\bibfnamefont{H.}~\bibnamefont{Jiang}},
  \bibinfo{author}{\bibfnamefont{H.}~\bibnamefont{Ding}},
  \bibinfo{author}{\bibfnamefont{M.}~\bibnamefont{Pu}}, \bibnamefont{and}
  \bibinfo{author}{\bibfnamefont{Z.}~\bibnamefont{Hou}}, \bibinfo{journal}{Soft
  matter} \textbf{\bibinfo{volume}{13}}, \bibinfo{pages}{836}
  (\bibinfo{year}{2017}).

\bibitem[{\citenamefont{Tailleur and Cates}(2008)}]{MIPS1}
\bibinfo{author}{\bibfnamefont{J.}~\bibnamefont{Tailleur}} \bibnamefont{and}
  \bibinfo{author}{\bibfnamefont{M.~E.} \bibnamefont{Cates}},
  \bibinfo{journal}{Phys. Rev. Lett.} \textbf{\bibinfo{volume}{100}},
  \bibinfo{pages}{218103} (\bibinfo{year}{2008}).

\bibitem[{\citenamefont{Redner et~al.}(2013{\natexlab{a}})\citenamefont{Redner,
  Hagan, and Baskaran}}]{MIPS2}
\bibinfo{author}{\bibfnamefont{G.~S.} \bibnamefont{Redner}},
  \bibinfo{author}{\bibfnamefont{M.~F.} \bibnamefont{Hagan}}, \bibnamefont{and}
  \bibinfo{author}{\bibfnamefont{A.}~\bibnamefont{Baskaran}},
  \bibinfo{journal}{Phys. Rev. Lett.} \textbf{\bibinfo{volume}{110}},
  \bibinfo{pages}{055701} (\bibinfo{year}{2013}{\natexlab{a}}).

\bibitem[{\citenamefont{Cates and Tailleur}(2015)}]{MIPS3}
\bibinfo{author}{\bibfnamefont{M.~E.} \bibnamefont{Cates}} \bibnamefont{and}
  \bibinfo{author}{\bibfnamefont{J.}~\bibnamefont{Tailleur}},
  \bibinfo{journal}{Annu. Rev. Condens. Matter Phys.}
  \textbf{\bibinfo{volume}{6}}, \bibinfo{pages}{219} (\bibinfo{year}{2015}).

\bibitem[{\citenamefont{Redner et~al.}(2013{\natexlab{b}})\citenamefont{Redner,
  Baskaran, and Hagan}}]{MIPS4}
\bibinfo{author}{\bibfnamefont{G.~S.} \bibnamefont{Redner}},
  \bibinfo{author}{\bibfnamefont{A.}~\bibnamefont{Baskaran}}, \bibnamefont{and}
  \bibinfo{author}{\bibfnamefont{M.~F.} \bibnamefont{Hagan}},
  \bibinfo{journal}{Phys. Rev. E} \textbf{\bibinfo{volume}{88}},
  \bibinfo{pages}{012305} (\bibinfo{year}{2013}{\natexlab{b}}).

\bibitem[{\citenamefont{Redner et~al.}(2016)\citenamefont{Redner, Wagner,
  Baskaran, and Hagan}}]{MIPS5}
\bibinfo{author}{\bibfnamefont{G.~S.} \bibnamefont{Redner}},
  \bibinfo{author}{\bibfnamefont{C.~G.} \bibnamefont{Wagner}},
  \bibinfo{author}{\bibfnamefont{A.}~\bibnamefont{Baskaran}}, \bibnamefont{and}
  \bibinfo{author}{\bibfnamefont{M.~F.} \bibnamefont{Hagan}},
  \bibinfo{journal}{Phys. Rev. Lett.} \textbf{\bibinfo{volume}{117}},
  \bibinfo{pages}{148002} (\bibinfo{year}{2016}).

\bibitem[{\citenamefont{Speck et~al.}(2014)\citenamefont{Speck, Bialk\'e,
  Menzel, and L\"owen}}]{MIPS6}
\bibinfo{author}{\bibfnamefont{T.}~\bibnamefont{Speck}},
  \bibinfo{author}{\bibfnamefont{J.}~\bibnamefont{Bialk\'e}},
  \bibinfo{author}{\bibfnamefont{A.~M.} \bibnamefont{Menzel}},
  \bibnamefont{and} \bibinfo{author}{\bibfnamefont{H.}~\bibnamefont{L\"owen}},
  \bibinfo{journal}{Phys. Rev. Lett.} \textbf{\bibinfo{volume}{112}},
  \bibinfo{pages}{218304} (\bibinfo{year}{2014}).

\bibitem[{\citenamefont{Speck et~al.}(2015)\citenamefont{Speck, Menzel,
  Bialk¨¦, and L\"owen}}]{MIPS7}
\bibinfo{author}{\bibfnamefont{T.}~\bibnamefont{Speck}},
  \bibinfo{author}{\bibfnamefont{A.~M.} \bibnamefont{Menzel}},
  \bibinfo{author}{\bibfnamefont{J.}~\bibnamefont{Bialk¨¦}}, \bibnamefont{and}
  \bibinfo{author}{\bibfnamefont{H.}~\bibnamefont{L\"owen}},
  \bibinfo{journal}{J. Chem. Phys.} \textbf{\bibinfo{volume}{142}},
  \bibinfo{pages}{224109} (\bibinfo{year}{2015}).

\bibitem[{\citenamefont{Bergmann et~al.}(2018)\citenamefont{Bergmann, Rapp, and
  Zimmermann}}]{MIPS8}
\bibinfo{author}{\bibfnamefont{F.}~\bibnamefont{Bergmann}},
  \bibinfo{author}{\bibfnamefont{L.}~\bibnamefont{Rapp}}, \bibnamefont{and}
  \bibinfo{author}{\bibfnamefont{W.}~\bibnamefont{Zimmermann}},
  \bibinfo{journal}{Phys. Rev. E} \textbf{\bibinfo{volume}{98}},
  \bibinfo{pages}{020603} (\bibinfo{year}{2018}).

\bibitem[{\citenamefont{Rapp et~al.}(2019)\citenamefont{Rapp, Bergmann, and
  Zimmermann}}]{MIPS9}
\bibinfo{author}{\bibfnamefont{L.}~\bibnamefont{Rapp}},
  \bibinfo{author}{\bibfnamefont{F.}~\bibnamefont{Bergmann}}, \bibnamefont{and}
  \bibinfo{author}{\bibfnamefont{W.}~\bibnamefont{Zimmermann}},
  \bibinfo{journal}{Eur. Phys. J. E} \textbf{\bibinfo{volume}{42}},
  \bibinfo{pages}{57} (\bibinfo{year}{2019}).

\bibitem[{\citenamefont{Takatori et~al.}(2014)\citenamefont{Takatori, Yan, and
  Brady}}]{MIPS12}
\bibinfo{author}{\bibfnamefont{S.~C.} \bibnamefont{Takatori}},
  \bibinfo{author}{\bibfnamefont{W.}~\bibnamefont{Yan}}, \bibnamefont{and}
  \bibinfo{author}{\bibfnamefont{J.~F.} \bibnamefont{Brady}},
  \bibinfo{journal}{Phys. Rev. Lett.} \textbf{\bibinfo{volume}{113}},
  \bibinfo{pages}{028103} (\bibinfo{year}{2014}).

\bibitem[{\citenamefont{Takatori and Brady}(2015)}]{MIPS13}
\bibinfo{author}{\bibfnamefont{S.~C.} \bibnamefont{Takatori}} \bibnamefont{and}
  \bibinfo{author}{\bibfnamefont{J.~F.} \bibnamefont{Brady}},
  \bibinfo{journal}{Phys. Rev. E} \textbf{\bibinfo{volume}{91}},
  \bibinfo{pages}{032117} (\bibinfo{year}{2015}).

\bibitem[{\citenamefont{Patch et~al.}(2017)\citenamefont{Patch, Yllanes, and
  Marchetti}}]{MIPS14}
\bibinfo{author}{\bibfnamefont{A.}~\bibnamefont{Patch}},
  \bibinfo{author}{\bibfnamefont{D.}~\bibnamefont{Yllanes}}, \bibnamefont{and}
  \bibinfo{author}{\bibfnamefont{M.~C.} \bibnamefont{Marchetti}},
  \bibinfo{journal}{Phys. Rev. E} \textbf{\bibinfo{volume}{95}},
  \bibinfo{pages}{012601} (\bibinfo{year}{2017}).

\bibitem[{\citenamefont{Wittkowski et~al.}(2014)\citenamefont{Wittkowski,
  Tiribocchi, Stenhammar, Allen, Marenduzzo, and Cates}}]{MIPS10}
\bibinfo{author}{\bibfnamefont{R.}~\bibnamefont{Wittkowski}},
  \bibinfo{author}{\bibfnamefont{A.}~\bibnamefont{Tiribocchi}},
  \bibinfo{author}{\bibfnamefont{J.}~\bibnamefont{Stenhammar}},
  \bibinfo{author}{\bibfnamefont{R.~J.} \bibnamefont{Allen}},
  \bibinfo{author}{\bibfnamefont{D.}~\bibnamefont{Marenduzzo}},
  \bibnamefont{and} \bibinfo{author}{\bibfnamefont{M.~E.} \bibnamefont{Cates}},
  \bibinfo{journal}{Nat. Commun.} \textbf{\bibinfo{volume}{5}},
  \bibinfo{pages}{4351} (\bibinfo{year}{2014}).

\bibitem[{\citenamefont{Tjhung et~al.}(2018)\citenamefont{Tjhung, Nardini, and
  Cates}}]{MIPS11}
\bibinfo{author}{\bibfnamefont{E.}~\bibnamefont{Tjhung}},
  \bibinfo{author}{\bibfnamefont{C.}~\bibnamefont{Nardini}}, \bibnamefont{and}
  \bibinfo{author}{\bibfnamefont{M.~E.} \bibnamefont{Cates}},
  \bibinfo{journal}{Phys. Rev. X} \textbf{\bibinfo{volume}{8}},
  \bibinfo{pages}{031080} (\bibinfo{year}{2018}).

\bibitem[{\citenamefont{Fily et~al.}(2014)\citenamefont{Fily, Henkes, and
  Marchetti}}]{MIPS15}
\bibinfo{author}{\bibfnamefont{Y.}~\bibnamefont{Fily}},
  \bibinfo{author}{\bibfnamefont{S.}~\bibnamefont{Henkes}}, \bibnamefont{and}
  \bibinfo{author}{\bibfnamefont{M.~C.} \bibnamefont{Marchetti}},
  \bibinfo{journal}{Soft Matter} \textbf{\bibinfo{volume}{10}},
  \bibinfo{pages}{2132} (\bibinfo{year}{2014}).

\bibitem[{\citenamefont{Z\"ottl and Stark}(2014)}]{MIPS16}
\bibinfo{author}{\bibfnamefont{A.}~\bibnamefont{Z\"ottl}} \bibnamefont{and}
  \bibinfo{author}{\bibfnamefont{H.}~\bibnamefont{Stark}},
  \bibinfo{journal}{Phys. Rev. Lett.} \textbf{\bibinfo{volume}{112}},
  \bibinfo{pages}{118101} (\bibinfo{year}{2014}).

\bibitem[{\citenamefont{Furukawa et~al.}(2014)\citenamefont{Furukawa,
  Marenduzzo, and Cates}}]{MIPS17}
\bibinfo{author}{\bibfnamefont{A.}~\bibnamefont{Furukawa}},
  \bibinfo{author}{\bibfnamefont{D.}~\bibnamefont{Marenduzzo}},
  \bibnamefont{and} \bibinfo{author}{\bibfnamefont{M.~E.} \bibnamefont{Cates}},
  \bibinfo{journal}{Phys. Rev. E} \textbf{\bibinfo{volume}{90}},
  \bibinfo{pages}{022303} (\bibinfo{year}{2014}).

\bibitem[{\citenamefont{Blaschke et~al.}(2016)\citenamefont{Blaschke, Maurer,
  Menon, Z?ttl, and Stark}}]{MIPS18}
\bibinfo{author}{\bibfnamefont{J.}~\bibnamefont{Blaschke}},
  \bibinfo{author}{\bibfnamefont{M.}~\bibnamefont{Maurer}},
  \bibinfo{author}{\bibfnamefont{K.}~\bibnamefont{Menon}},
  \bibinfo{author}{\bibfnamefont{A.}~\bibnamefont{Z?ttl}}, \bibnamefont{and}
  \bibinfo{author}{\bibfnamefont{H.}~\bibnamefont{Stark}},
  \bibinfo{journal}{Soft Matter} \textbf{\bibinfo{volume}{12}},
  \bibinfo{pages}{9821} (\bibinfo{year}{2016}).

\bibitem[{\citenamefont{Stenhammar et~al.}(2015)\citenamefont{Stenhammar,
  Wittkowski, Marenduzzo, and Cates}}]{MIPS19}
\bibinfo{author}{\bibfnamefont{J.}~\bibnamefont{Stenhammar}},
  \bibinfo{author}{\bibfnamefont{R.}~\bibnamefont{Wittkowski}},
  \bibinfo{author}{\bibfnamefont{D.}~\bibnamefont{Marenduzzo}},
  \bibnamefont{and} \bibinfo{author}{\bibfnamefont{M.~E.} \bibnamefont{Cates}},
  \bibinfo{journal}{Phys. Rev. Lett.} \textbf{\bibinfo{volume}{114}},
  \bibinfo{pages}{018301} (\bibinfo{year}{2015}).

\bibitem[{\citenamefont{Dolai et~al.}(2018)\citenamefont{Dolai, Simha, and
  Mishra}}]{MIPS20}
\bibinfo{author}{\bibfnamefont{P.}~\bibnamefont{Dolai}},
  \bibinfo{author}{\bibfnamefont{A.}~\bibnamefont{Simha}}, \bibnamefont{and}
  \bibinfo{author}{\bibfnamefont{S.}~\bibnamefont{Mishra}},
  \bibinfo{journal}{Soft Matter} \textbf{\bibinfo{volume}{14}},
  \bibinfo{pages}{6137} (\bibinfo{year}{2018}).

\bibitem[{\citenamefont{Rogel~Rodriguez
  et~al.}(2020)\citenamefont{Rogel~Rodriguez, Alarcon, Martinez, Ram¨ªrez, and
  Valeriani}}]{MIPS21}
\bibinfo{author}{\bibfnamefont{D.}~\bibnamefont{Rogel~Rodriguez}},
  \bibinfo{author}{\bibfnamefont{F.}~\bibnamefont{Alarcon}},
  \bibinfo{author}{\bibfnamefont{R.}~\bibnamefont{Martinez}},
  \bibinfo{author}{\bibfnamefont{J.}~\bibnamefont{Ram¨ªrez}}, \bibnamefont{and}
  \bibinfo{author}{\bibfnamefont{C.}~\bibnamefont{Valeriani}},
  \bibinfo{journal}{Soft Matter} \textbf{\bibinfo{volume}{16}},
  \bibinfo{pages}{1162} (\bibinfo{year}{2020}).

\bibitem[{\citenamefont{Stenhammar et~al.}(2014)\citenamefont{Stenhammar,
  Marenduzzo, Allen, and Cates}}]{MIPS22}
\bibinfo{author}{\bibfnamefont{J.}~\bibnamefont{Stenhammar}},
  \bibinfo{author}{\bibfnamefont{D.}~\bibnamefont{Marenduzzo}},
  \bibinfo{author}{\bibfnamefont{R.~J.} \bibnamefont{Allen}}, \bibnamefont{and}
  \bibinfo{author}{\bibfnamefont{M.~E.} \bibnamefont{Cates}},
  \bibinfo{journal}{Soft Matter} \textbf{\bibinfo{volume}{10}},
  \bibinfo{pages}{1489} (\bibinfo{year}{2014}).

\bibitem[{\citenamefont{Siebert et~al.}(2017)\citenamefont{Siebert, Letz,
  Speck, and Virnau}}]{MIPS23}
\bibinfo{author}{\bibfnamefont{J.~T.} \bibnamefont{Siebert}},
  \bibinfo{author}{\bibfnamefont{J.}~\bibnamefont{Letz}},
  \bibinfo{author}{\bibfnamefont{T.}~\bibnamefont{Speck}}, \bibnamefont{and}
  \bibinfo{author}{\bibfnamefont{P.}~\bibnamefont{Virnau}},
  \bibinfo{journal}{Soft Matter} \textbf{\bibinfo{volume}{13}},
  \bibinfo{pages}{1020} (\bibinfo{year}{2017}).

\bibitem[{\citenamefont{Liao and Klapp}(2018)}]{MIPS24}
\bibinfo{author}{\bibfnamefont{G.-J.} \bibnamefont{Liao}} \bibnamefont{and}
  \bibinfo{author}{\bibfnamefont{S.~H.~L.} \bibnamefont{Klapp}},
  \bibinfo{journal}{Soft Matter} \textbf{\bibinfo{volume}{14}},
  \bibinfo{pages}{7873} (\bibinfo{year}{2018}).

\bibitem[{\citenamefont{Celani et~al.}(2012)\citenamefont{Celani, Bo, Eichhorn,
  and Aurell}}]{inertia12}
\bibinfo{author}{\bibfnamefont{A.}~\bibnamefont{Celani}},
  \bibinfo{author}{\bibfnamefont{S.}~\bibnamefont{Bo}},
  \bibinfo{author}{\bibfnamefont{R.}~\bibnamefont{Eichhorn}}, \bibnamefont{and}
  \bibinfo{author}{\bibfnamefont{E.}~\bibnamefont{Aurell}},
  \bibinfo{journal}{Phys. Rev. Lett.} \textbf{\bibinfo{volume}{109}},
  \bibinfo{pages}{260603} (\bibinfo{year}{2012}).

\bibitem[{\citenamefont{Shankar and Marchetti}(2018)}]{inertia13}
\bibinfo{author}{\bibfnamefont{S.}~\bibnamefont{Shankar}} \bibnamefont{and}
  \bibinfo{author}{\bibfnamefont{M.~C.} \bibnamefont{Marchetti}},
  \bibinfo{journal}{Phys. Rev. E} \textbf{\bibinfo{volume}{98}},
  \bibinfo{pages}{020604} (\bibinfo{year}{2018}).

\bibitem[{\citenamefont{Crosato et~al.}(2019)\citenamefont{Crosato, Prokopenko,
  and Spinney}}]{inertia14}
\bibinfo{author}{\bibfnamefont{E.}~\bibnamefont{Crosato}},
  \bibinfo{author}{\bibfnamefont{M.}~\bibnamefont{Prokopenko}},
  \bibnamefont{and} \bibinfo{author}{\bibfnamefont{R.~E.}
  \bibnamefont{Spinney}}, \bibinfo{journal}{Phys. Rev. E}
  \textbf{\bibinfo{volume}{100}}, \bibinfo{pages}{042613}
  (\bibinfo{year}{2019}).

\bibitem[{\citenamefont{Mandal et~al.}(2019)\citenamefont{Mandal, Liebchen, and
  L\"owen}}]{inertia10}
\bibinfo{author}{\bibfnamefont{S.}~\bibnamefont{Mandal}},
  \bibinfo{author}{\bibfnamefont{B.}~\bibnamefont{Liebchen}}, \bibnamefont{and}
  \bibinfo{author}{\bibfnamefont{H.}~\bibnamefont{L\"owen}},
  \bibinfo{journal}{Phys. Rev. Lett.} \textbf{\bibinfo{volume}{123}},
  \bibinfo{pages}{228001} (\bibinfo{year}{2019}).

\bibitem[{\citenamefont{L{\"o}wen}(2020)}]{inertia1}
\bibinfo{author}{\bibfnamefont{H.}~\bibnamefont{L{\"o}wen}},
  \bibinfo{journal}{J. Chem. Phys.} \textbf{\bibinfo{volume}{152}},
  \bibinfo{pages}{040901} (\bibinfo{year}{2020}).

\bibitem[{\citenamefont{Dai et~al.}(2020)\citenamefont{Dai, Bruss, and
  Glotzer}}]{inertia11}
\bibinfo{author}{\bibfnamefont{C.}~\bibnamefont{Dai}},
  \bibinfo{author}{\bibfnamefont{I.~R.} \bibnamefont{Bruss}}, \bibnamefont{and}
  \bibinfo{author}{\bibfnamefont{S.~C.} \bibnamefont{Glotzer}},
  \bibinfo{journal}{Soft Matter} \textbf{\bibinfo{volume}{16}},
  \bibinfo{pages}{2847} (\bibinfo{year}{2020}).

\bibitem[{\citenamefont{Walsh et~al.}(2017)\citenamefont{Walsh, Wagner,
  Schlossberg, Olson, Baskaran, and Menon}}]{inertia4}
\bibinfo{author}{\bibfnamefont{L.}~\bibnamefont{Walsh}},
  \bibinfo{author}{\bibfnamefont{C.~G.} \bibnamefont{Wagner}},
  \bibinfo{author}{\bibfnamefont{S.}~\bibnamefont{Schlossberg}},
  \bibinfo{author}{\bibfnamefont{C.}~\bibnamefont{Olson}},
  \bibinfo{author}{\bibfnamefont{A.}~\bibnamefont{Baskaran}}, \bibnamefont{and}
  \bibinfo{author}{\bibfnamefont{N.}~\bibnamefont{Menon}},
  \bibinfo{journal}{Soft Matter} \textbf{\bibinfo{volume}{13}},
  \bibinfo{pages}{8964} (\bibinfo{year}{2017}).

\bibitem[{\citenamefont{Dauchot and D\'emery}(2019)}]{inertia5}
\bibinfo{author}{\bibfnamefont{O.}~\bibnamefont{Dauchot}} \bibnamefont{and}
  \bibinfo{author}{\bibfnamefont{V.}~\bibnamefont{D\'emery}},
  \bibinfo{journal}{Phys. Rev. Lett.} \textbf{\bibinfo{volume}{122}},
  \bibinfo{pages}{068002} (\bibinfo{year}{2019}).

\bibitem[{\citenamefont{Valani et~al.}(2019)\citenamefont{Valani, Slim, and
  Simula}}]{inertia6}
\bibinfo{author}{\bibfnamefont{R.~N.} \bibnamefont{Valani}},
  \bibinfo{author}{\bibfnamefont{A.~C.} \bibnamefont{Slim}}, \bibnamefont{and}
  \bibinfo{author}{\bibfnamefont{T.}~\bibnamefont{Simula}},
  \bibinfo{journal}{Phys. Rev. Lett.} \textbf{\bibinfo{volume}{123}},
  \bibinfo{pages}{024503} (\bibinfo{year}{2019}).

\bibitem[{\citenamefont{Couder and Fort}(2006)}]{inertia7}
\bibinfo{author}{\bibfnamefont{Y.}~\bibnamefont{Couder}} \bibnamefont{and}
  \bibinfo{author}{\bibfnamefont{E.}~\bibnamefont{Fort}},
  \bibinfo{journal}{Phys. Rev. Lett.} \textbf{\bibinfo{volume}{97}},
  \bibinfo{pages}{154101} (\bibinfo{year}{2006}).

\bibitem[{\citenamefont{Rabault et~al.}(2019)\citenamefont{Rabault, Fauli, and
  Carlson}}]{inertia8}
\bibinfo{author}{\bibfnamefont{J.}~\bibnamefont{Rabault}},
  \bibinfo{author}{\bibfnamefont{R.~A.} \bibnamefont{Fauli}}, \bibnamefont{and}
  \bibinfo{author}{\bibfnamefont{A.}~\bibnamefont{Carlson}},
  \bibinfo{journal}{Phys. Rev. Lett.} \textbf{\bibinfo{volume}{122}},
  \bibinfo{pages}{024501} (\bibinfo{year}{2019}).

\bibitem[{\citenamefont{Ivlev et~al.}(2015)\citenamefont{Ivlev, Bartnick,
  Heinen, Du, Nosenko, and L\"owen}}]{inertia9}
\bibinfo{author}{\bibfnamefont{A.~V.} \bibnamefont{Ivlev}},
  \bibinfo{author}{\bibfnamefont{J.}~\bibnamefont{Bartnick}},
  \bibinfo{author}{\bibfnamefont{M.}~\bibnamefont{Heinen}},
  \bibinfo{author}{\bibfnamefont{C.-R.} \bibnamefont{Du}},
  \bibinfo{author}{\bibfnamefont{V.}~\bibnamefont{Nosenko}}, \bibnamefont{and}
  \bibinfo{author}{\bibfnamefont{H.}~\bibnamefont{L\"owen}},
  \bibinfo{journal}{Phys. Rev. X} \textbf{\bibinfo{volume}{5}},
  \bibinfo{pages}{011035} (\bibinfo{year}{2015}).

\bibitem[{\citenamefont{Scholz et~al.}(2018)\citenamefont{Scholz, Jahanshahi,
  Ldov, and L{\"o}wen}}]{inertia3}
\bibinfo{author}{\bibfnamefont{C.}~\bibnamefont{Scholz}},
  \bibinfo{author}{\bibfnamefont{S.}~\bibnamefont{Jahanshahi}},
  \bibinfo{author}{\bibfnamefont{A.}~\bibnamefont{Ldov}}, \bibnamefont{and}
  \bibinfo{author}{\bibfnamefont{H.}~\bibnamefont{L{\"o}wen}},
  \bibinfo{journal}{Nat. Commun.} \textbf{\bibinfo{volume}{9}},
  \bibinfo{pages}{5156} (\bibinfo{year}{2018}).

\bibitem[{\citenamefont{Leyman et~al.}(2018)\citenamefont{Leyman, Ogemark,
  Wehr, and Volpe}}]{inertia2}
\bibinfo{author}{\bibfnamefont{M.}~\bibnamefont{Leyman}},
  \bibinfo{author}{\bibfnamefont{F.}~\bibnamefont{Ogemark}},
  \bibinfo{author}{\bibfnamefont{J.}~\bibnamefont{Wehr}}, \bibnamefont{and}
  \bibinfo{author}{\bibfnamefont{G.}~\bibnamefont{Volpe}},
  \bibinfo{journal}{Phys. Rev. E} \textbf{\bibinfo{volume}{98}},
  \bibinfo{pages}{052606} (\bibinfo{year}{2018}).

\end{thebibliography}

\end{document}